\documentclass[doublecol]{epl2}
\def\maj#1{\ifmmode\mbox{\usefont{U}{msb}{m}{n}#1}\else{\usefont{U}{msb}{m}{n}#1}\fi}
\def\v#1{\mathbf{#1}}
\usepackage{graphics,epsfig,graphicx}
\usepackage{color}

\title{\textbf{Analytical approach to semiconductor Bloch equations}}
\author{Monique Combescot\inst{1}, Odile Betbeder-Matibet\inst{1} and Michael N. Leuenberger\inst{2}}
\institute{
\inst{1} Institut des NanoSciences de Paris,
Universit\'e Pierre et Marie Curie, 
CNRS, Campus Boucicaut, 140 rue de Lourmel, 75015 Paris\\
\inst{2} NanoScience Technology Center and Dept.\ of Physics, University of Central Florida,
12424 Research Parkway Suite 400, Orlando, FL 32826, USA}

\abstract{
Although semiconductor Bloch equations have been widely used for decades to address ultrafast optical phenomena in semiconductors, they have a few important drawbacks: (i) Coulomb terms between free electron-hole pairs require Hartree-Fock treatment which, in its usual form, preserves excitonic poles but loses biexcitonic resonances. (ii) Solving the resulting coupled differential equations imposes heavy numerics which completely hide the physics. This can be completely avoided if, instead of free electron-hole pairs, we use correlated pairs, i.e., excitons. Their interactions are easy to handle through the recently constructed composite-exciton many-body theory, which allows us to \emph{analytically} obtain the time evolution of the polarization induced by a laser pulse. This polarization comes from Coulomb interactions between virtual excitons, but also from Coulomb-free fermion exchanges, which are dominant at large detuning.}

\begin{document}

\maketitle

\section{Introduction}

Interaction of semiconductors with ultrafast laser pulses has been widely studied for decades. Light-matter interaction leads to a large variety of nonlinear effects of great interest in semiconductor technology. Probably the most fascinating ones are those induced by unabsorbed photons: while these photons could be thought of being completely inactive, they do act on semiconductors in a quite subtle manner through their coupling to virtual excitons. These virtual excitons interact with carriers, either electrons or holes already present in the sample, not only through the Coulomb potential, but mostly through the Pauli exclusion principle. The many-body theory for composite excitons we have recently constructed \cite{Phys.Rep.} makes this Pauli exclusion principle appear by means of  ``Pauli scatterings'' between excitons. They correspond to fermion exchange in the absence of fermion interaction, as  nicely visualized by the so-called Shiva diagrams \cite{Shiva}. These Pauli scatterings in fact control all nonlinear optical effects in semiconductors. Among those induced by unabsorbed photons, we can cite the exciton optical Stark shift \cite{Stark}, Faraday rotation \cite{Faraday} in photoexcited semiconductors, precession \cite{Precession} and teleportation \cite{Leuenberger_teleportation,Teleportation,Leuenberger_book} of trapped electron spin, quantum computing \cite{Leuenberger_QC,Seigneur,Leuenberger_book} and classical optical computing \cite{Thompson} using trapped spins. Effects induced by these virtual excitons decrease with photon detuning but increase with photon number, i.e., pulse intensity.

From a theoretical point of view, these effects face a major problem. As excitons play a key role, we must include electron-hole Coulomb interaction exactly, in order to possibly reach their associated poles. However, many-body effects are known to be approached through perturbative expansion only. This is fully reasonable for interaction between the exciton fermionic components, since excitons are neutral objects, so that their Coulomb interaction should induce small effects. Nevertheless, an exact treatment of the electron-hole interaction is required to get the excitonic poles. The clean separation between electron-hole ladder processes binding the exciton and those entering exciton-exciton interaction has stayed an open problem for decades \cite{K-K}. One of the advantages of the composite-exciton many-body theory  \cite{Phys.Rep.} is to provide such a separation.

In addition to the difficulty linked to the electron-hole interaction, excitons also have an intrinsic difficulty linked to their compositeness, which leads to a serious ambiguity when trying to pair up electrons and holes to form excitons. We have shown \cite{Phys.Rep.} that, as a direct consequence of the Pauli exclusion principle, it is impossible to properly handle the interaction between excitons through an effective exciton Hamiltonian reading as $H_{X}+V_{XX}$. The potential $V_{XX}$, which describes interactions between excitons considered as elementary bosons, contains exciton-exciton scatterings. These are derived from the Coulomb scatterings between elementary carriers through bosonization procedures which include some fermion exchanges. With the Hamiltonian not written as  $H_{X}+V_{XX}$, it is not possible to use standard many-body techniques since they all rely on perturbative expansion in interaction potential. A totally different path had to be constructed to handle many-body effects with excitons. The many-body theory we proposed \cite{Phys.Rep.} is quite original as it uses an operator algebra, instead of the scalar algebra used in Green's functions. The idea is to include the major difficulty of the problem, namely, the exact summation of all electron-hole processes giving rise to excitonic poles, in the very first line of the theory, by working with exciton operators. These operators have an important drawback since excitons are composite bosons with commutation relations not as strict as for elementary particles. We can however deal with this difficulty by introducing ``Pauli scatterings'' $\lambda\left(^{n\ \,j}_{m\ i}\right)$ which describe fermion exchanges in the absence of fermion interaction. They appear in
\begin{equation}
[B_m,B_i^\dag]=\delta_{m,i}-D_{mi}\ ,
\end{equation}
\begin{equation}
[D_{mi},B_j^\dag]=\sum_n\left[\lambda\left(^{n\ \,j}_{m\ i}\right)+(m\leftrightarrow n)\right]\,B_n^\dag\ .
\end{equation}

Since exciton operators $B_i^\dag$ contain the tricky electron-hole part of the Coulomb interaction leading to excitonic poles in an exact way, the rest of Coulomb interaction can be treated perturbatively through exciton-exciton Coulomb scatterings $\xi\left(^{n\ \,j}_{m\ i}\right)$. They appear in
\begin{equation}
[H,B_i^\dag]=E_i\,B_i^\dag+V_i^\dag\ ,
\end{equation}
\begin{equation}
[V_i^\dag,B_j^\dag]=\sum_{mn}\xi\left(^{n\ \,j}_{m\ i}\right)B_m^\dag B_n^\dag\ .
\end{equation}
These Pauli and Coulomb scatterings allow us to calculate any physical quantity dealing with excitons in an easy way.

Before using this new many-body theory to calculate the time evolution of the polarization induced by a laser pulse, we wish to emphasize that the semiconductor Bloch equations (SBE) \cite{Haug1,Binder1,Lindberg1,Lindberg2,Axt,Binder2,Haug2,Erem} use free pairs $a_{\v k_e}^\dag b_{\v k_h}^\dag$ instead of excitons $B_i^\dag$. This just corresponds to a basis change, since these operators are linked by 
\begin{equation}
B_i^\dag=\sum_{\v k_e,\v k_h} a_{\v k_e}^\dag b_{\v k_h}^\dag\,\langle\v k_h,\v k_e|i\rangle\ ,
\end{equation}
\begin{equation}
a_{\v k_e}^\dag b_{\v k_h}^\dag=\sum_iB_i^\dag\,\langle i|\v k_e,\v k_h\rangle\ .
\end{equation}

Difficulties with SBE actually come from using these free-pair operators in problems in which excitons play a key role.
We here show, by working with exciton operators in the form $\langle B_p^\dag\rangle_t$ instead of $\langle a_{\v k}^\dag b_{-\v k}^\dag\rangle_t$, that we can calculate such an expectation value analytically, thereby avoiding the problem of solving numerically the coupled differential equations. This is definitely a great advantage because the physical meaning of the various contributions to the polarization then appear in a transparent way. We find, once more,  that one part of the polarization comes from Coulomb interaction between excitons while the other part comes from the Pauli exclusion principle between the exciton fermionic components, which, as usual, is dominant for large photon detuning.

\section{Proposed approach}

We calculate the polarization induced by a single laser pulse starting from the non-excited semiconductor. Theoretical extensions of this approach to semiconductors already containing carriers or to experiments with more than one type of photons, such as in four-wave mixing, can be done along the same line. These will be presented elsewhere, as the goal of this letter is to present the main steps of the new procedure we propose.

The time evolution of the exciton polarization can be written in terms of $\langle\psi_t|B_p^\dag|\psi_t\rangle=\langle B_p^\dag\rangle_t$, where $|\psi_t\rangle$ is the semiconductor state resulting from its coupling $W_t$ to the photon field. This mean value obeys
\begin{equation}
-i\,\frac{\partial}{\partial t}\langle B_p^\dag\rangle_t=\langle[H_{sc}+W_t,B_p^\dag]\rangle_t\ .
\end{equation}
$H_{sc}$ is the semiconductor Hamiltonian in the absence of photon, while 
$W_t=U_t+U_t^\dag$, with $U_t=\sum_i\Omega_i (t)B_i$, the time dependence of the Rabi coupling $\Omega_i(t)$ to exciton $i$ following the laser pulse. Since for composite-boson operators $[B_m^\dag,B_i^\dag]=0$, so that $[U_t^\dag,B_p^\dag]=0$, the commutation relations in Eqs.(1,3) readily turn Eq.(7) into
\begin{eqnarray}
\left(-i\frac{\partial}{\partial t}-E_p\right)\langle B_p^\dag\rangle_t=\Omega_p(t)\hspace{3cm}\nonumber\\
 -\sum_i\Omega_i (t)\langle D_{ip}\rangle_t+\langle V_p^\dag\rangle_t \ .
\end{eqnarray}
Beside a bare linear term $\Omega_p (t)$ in photon coupling, the time evolution of 
$\langle B_p^\dag\rangle_t$ contains contributions coming from the composite nature of exciton $p$ through $D_{ip}$, and from Coulomb interaction with exciton $p$ through $V_p^\dag$.

If we only keep resonant terms, for a laser field with frequency $\omega$ introduced adiabatically from $t=-\infty$ with a $1/\epsilon$ risetime, the time dependence of the Rabi coupling $\Omega_i (t)$ reads $\Omega_i e^{i\omega t}
e^{\epsilon t}$, so that, for a non-excited semiconductor, $|\psi_{t=-\infty}\rangle=|v\rangle$,
the linear term $\Omega_p (t)$ in Eq.(8) gives the free part of $\langle B_p^\dag\rangle_t$ as $\Omega_p e^{i\omega t}e^{\epsilon t}/(\omega-E_p-i\epsilon)$. 

However, even if adiabatic establishment is considered in most papers, it does not seem to us the proper way to describe ultrashort laser pulses with steep rise time and duration time $T_p$ long compared to photon period $\omega^{-1}$. 
We find it more appropriate to take $\Omega_i (t)$ as $\Omega_i e^{i\omega t}$ during the pulse duration, $0<t<T_p$ and zero otherwise, with $|\psi_t\rangle=|v\rangle$ for $t\leq0_+$ (sudden approximation, valid for steep rise time). The free part of $\langle B_p^\dag\rangle_t$, which now cancels for $t=0$ instead of $t=-\infty$, reads as
\begin{equation}
e^{-i\omega t}\langle B_p^\dag\rangle_t^\mathrm{free}=e^{-i\omega_p t}\Omega_p\,\Delta_t^{(1)}(\omega_p)
\end{equation}
where $\omega_p=\omega-E_p$ is the photon detuning with respect to exciton $p$, while $\Delta_t^{(1)}(x)=(e^{ixt}-1)/x$, so that $\Delta_t^{(1)}(0)=it$. Note that, as $\Delta_t^{(1)}(x)$ also reads $2i\pi e^{ixt/2}\delta_t(x)$, where $\delta_t(x)=(\pi x)^{-1}
\sin(xt/2)$ is a ``delta function'' of width $2/t$, this free part barely leads to the Fermi golden rule. 

In addition to its physical relevance, a sudden rise for $\Omega_i (t)$ allows us to eliminate $t$ from the Hamiltonian through the so-called ``rotating frame'' unitary transformation that we can take \cite{unit.transf.} as
$Z_t=\exp (-i\omega t \sum a_{\v k}^\dag
a_{\v k})$. This gives $Z_t^{-1}a_{\v k}^\dag Z_t=e^{i\omega t}a_{\v k}^\dag$, while $b_{\v k}^\dag$ stays unchanged. The effective Hamiltonian $\tilde{H}=Z_t^{-1}H_tZ_t-
iZ_t^{-1}\dot{Z}_t$, which rules the time evolution of $|\tilde{\psi}_t\rangle=Z_t^{-1}|\psi_t\rangle$, reduces to $\tilde{H}_{sc}+\tilde{W}$, where $\tilde{H}_{sc}$ is just $H_{sc}$ with all electron energies shifted by $-\omega$, which amounts to replace $E_i$ by $E_i-\omega$, with the coupling $\tilde{W}$ reading as $\tilde{U}+\tilde{U}^\dag$ where $\tilde{U}=\sum_i\Omega_i B_i$ is now $t$-independent. The time evolution of
\begin{equation}
 \langle\tilde{\psi}_t|B_p^\dag|\tilde{\psi}_t\rangle=\langle\langle B_p^\dag\rangle\rangle_t=e^{-i\omega t}\langle B_p^\dag\rangle_t
\end{equation}
follows from Eq.(8), with $-E_p$ replaced by $\omega_p$ and $\Omega_i(t)$ by $\Omega_i$.

The linear term $\Omega_p$ leads to $\langle\langle B_p^\dag\rangle\rangle_t^\mathrm{free}$, given in Eq.(9). To get Pauli and Coulomb contributions  coming from $\langle\langle D_{ip}\rangle\rangle_t$ and $\langle\langle V_p^\dag\rangle\rangle_t$, we note that,
since $\tilde{H}$ is now time-independent, we then have $|\tilde{\psi}_t\rangle=e^{-i\tilde
{H}t}|v\rangle$, in which we can use the integral representation of the exponential,
\begin{equation}
e^{-i\tilde{H}t}=\int_{-\infty}^{+\infty}\frac{dx}{(-2i\pi)}\,\frac{e^{-i(x+i0_+)t}}{x+i0_+-\tilde{H}}\ ,
\end{equation}
and decouple interaction with photons from interactions between carriers, through
\begin{eqnarray}
\frac{1}{a-\tilde{H}}&=&\frac{1}{a-\tilde{H}_{sc}}+\frac{1}{a-\tilde{H}}\,\tilde{W}\,\frac{1}{a-\tilde{H}_{sc}}\nonumber\\
&=&\sum_{n=0}^{+\infty}\left(\frac{1}{a-\tilde{H}_{sc}}\,\tilde{W}\right)^n\,\frac{1}{a-\tilde{H}_{sc}}\ .
\end{eqnarray}

As $\tilde{W}$ changes the number of pairs by one while $D_{ip}$ conserves this number, $ \langle\langle D_{ip}\rangle\rangle_t$ only has even terms in photon coupling, while $\langle\langle V_p^\dag\rangle\rangle_t$ only has odd terms since $V_p^\dag$ creates a pair. Therefore, as expected, the polarization given in Eq.(8) has odd terms only in Rabi coupling $\Omega_i$.

Eq.(12) generates the polarization as an expansion in photon interaction $\tilde{W}$. Such an expansion is a priori valid for small laser intensity. We wish to stress that this limitation is fully consistent with  only considering $\langle\langle B\rangle\rangle_t$ but not $\langle\langle B^n\rangle\rangle_t$ with $n\geq 2$. In the case of large photon field for which such a $\tilde{W}$ expansion is not valid, the relevant operators are no more exciton, but polariton operators. In a near future, we will address large photon fields using our work on interacting polaritons \cite{Pol1,Pol2}, along ideas similar to the ones we propose here. However, polaritons being far more complex composite bosons than excitons, this extension is at the present time beyond our scope.

\section{Pauli part of $\langle\langle B_p^\dag\rangle\rangle_t$}

As shown in the appendix, the Pauli part of $\langle\langle B_p^\dag\rangle\rangle_t$, which comes from  the second term of Eq.(8), has a third order contribution in photon coupling which reads
\begin{equation}
\langle\langle B_p^\dag\rangle\rangle_t^\mathrm{Pauli}\simeq 2e^{-i\omega_pt}\sum_{ijk}\Omega_i
\Omega_j\Omega_k^\ast\,\lambda_{ijkp}\Delta_t^{(3)}(\omega_p,-\omega_j,\omega_k)\ ,
\end{equation}
where $2\lambda_{ijkp}\equiv\left[\lambda\left(^{j\ k}_{i\ p}\right)+(i\leftrightarrow j)\right]$ while the $\Delta_t^{(n)}$ functions are linked by
\begin{eqnarray}
x_{n+1}\Delta_t^{(n+1)}(x_1,x_2,\cdots,x_{n+1})=\hspace{2cm}\nonumber\\
\Delta_t^{(n)}(x_1+x_{n+1},x_2,\cdots,x_n)-\Delta_t^{(n)}(x_1,x_2,\cdots,x_n)\ .
\end{eqnarray}

\section{Coulomb part of $\langle\langle B_p^\dag\rangle\rangle_t$}

The Coulomb part of $\langle\langle B_p^\dag\rangle\rangle_t$, comes from the last term of Eq.(8). In contrast with $\langle\langle D_{ip}\rangle\rangle_t$, which only reads in terms of exciton energies and wave functions, the exact calculation of $\langle\langle V_p^\dag\rangle\rangle_t$ at third order in photon coupling requires the knowledge of the whole two-pair eigenstate spectrum. We can however approximate this Coulomb part in two limits: 

(i) For materials having a well-separated biexciton, the Coulomb part of the polarization for photons close to the biexciton resonance is found to read
\begin{eqnarray}
\langle\langle B_p^\dag\rangle\rangle_t^\mathrm{biexc}\simeq -e^{-i\omega_p t}\sum_{ijk}\Omega_i
\Omega_j\Omega_k \xi_{ijkp}^{(XX)}\nonumber\\
\times\ \Delta_t^{(4)}(\omega_p,E_{XX}-2\omega+\omega_j,-\omega_j,\omega_k)\ ;
\end{eqnarray}
$\xi_{ijkp}^{(XX)}$ can be seen as a Coulomb interaction between excitons mediated by the molecular biexciton resonance,
\begin{equation}
2\xi_{ijkp}^{(XX)}=\sum_{m,n}\langle v|B_iB_j|XX\rangle\langle XX|B_m^\dag B_n^\dag|v\rangle
\xi\left(^{n\ \,k}_{m\ p}\right)\ ,
\end{equation}
where $|XX\rangle$ and $E_{XX}$ are the molecular biexciton state and energy, these being obtained through numerics only.

(ii) We can also expand the Coulomb part of the polarization in Coulomb scattering divided by detuning, through 
\begin{equation}
\frac{1}{a-\tilde{H}_{sc}}\, B_j^\dag=\left(B_j^\dag+\frac{1}{a-\tilde{H}_{sc}}\,V_j^\dag\right)\frac{1}
{a-\tilde{H}_{sc}+\omega_j}\ .
\end{equation}
As shown in the appendix, we then find
\begin{eqnarray}
\langle\langle B_p^\dag\rangle\rangle_t^\mathrm{Coulomb}\simeq -\,e^{-i\omega_pt}\sum_{ijk} \Omega_i
\Omega_j\Omega_k^\ast\,\hat{\xi}_{ijkp}\nonumber\\
\times\ \Delta_t^{(4)}(\omega_p,-\omega_i,-\omega_j,\omega_k))\ ,
\end{eqnarray}
where $\hat{\xi}$ is the mixed
direct-exchange Coulomb scattering standard for time evolution \cite{Phys.Rep.},
\begin{equation}
2\hat{\xi}_{ijkp}=\left[\xi\left(^{j\ k}_{i\ p}\right)-\xi^\mathrm{in}\left(^{j\ k}_{i\ p}\right)\right]+(i\leftrightarrow
j)\ ,
\end{equation}
the ``in'' Coulomb exchange part of this scattering being equal to $\sum_{m,n}\lambda\left(^{j\ \,n}_{i\ m}\right)\xi
\left(^{n\ \,k}_{m\ p}\right)$.

\section{Discussion}

The above derivation shows that $\langle\langle B_p^\dag\rangle\rangle_t$ contains two cubic terms in photon coupling given in Eq.(13) and Eq.(18) --- or Eq.(15) when the molecular biexciton plays a key role. A term comes from fermion exchange with exciton $p$ without Coulomb process, the other from Coulomb interactions with exciton $p$ with or without fermion exchange. They basically have the same structure except that, Coulomb scatterings $\xi_{ijkp}$ being energylike quantities while Pauli scatterings $\lambda_{ijkp}$ are dimensionless, the Coulomb part of the polarization has one more detuning in the denominator than the Pauli part: indeed, $\Delta_t^{(3)}$ has $\omega_j\omega_k$, while $\Delta_t^{(4)}$ has $\omega_i\omega_j\omega_k$. Consequently, as usual for nonlinear optical effects, the large detuning behavior of $\langle\langle B_p^\dag\rangle\rangle_t$ is fully controlled by the Pauli exclusion principle with exciton $p$ in the absence of Coulomb interaction.

Furthermore, we note that the coupling $\Omega_i$ to exciton $i$ depends on $i$ through $\langle \v r=\v 0|\nu_i\rangle$, where $\langle\v r|\nu_i\rangle$ is the $i$ exciton relative motion wave function. This makes photons dominantly coupled to $s$ excitons. In addition, $\langle\langle B_p^\dag\rangle\rangle_t$ decreases when photon detuning increases. Therefore, sizeable
$\langle\langle B_p^\dag\rangle\rangle_t$ comes from photon detuning to the $1s$ exciton ground state (labelled 0) not too large. For such photons, the $\omega_i\omega_j\omega_k$ denominator of $\Delta_t^{(4)}$ in Eq.(18) leads us to only keep $(i,j,k)=0$ in the sum. Similarly, the $\omega_j\omega_k$ denominator of $\Delta_t^{(3)}$ in Eq.(13) leads us to only keep
$(j,k)=0$. The cubic term of $\langle B_p^\dag\rangle_t$ then appears as 
\begin{eqnarray}
\langle B_p^\dag\rangle_t^{(3)}\simeq e^{iE_pt}\Omega_0|\Omega_0|^2\left[2\hat{\lambda}_{000p}\Delta_t^{(3)}(\omega_p,-\omega_0,\omega_0)
\right.\nonumber\\
\left.-\hat{\xi}_{000p}\Delta_t^{(4)}(\omega_p,-\omega_0,-\omega_0,\omega_0)\right]\ ,
\end{eqnarray}
where $\Omega_0\hat{\lambda}_{000p}=\sum_i\Omega_i\lambda_{i00p}$, the summation over $i$ being performed through closure relation. The above result is valid for $\omega_0=\omega-E_0$ small compared to $\omega_{i\neq 0}$ but still large enough to possibly keep one term only in the $\langle\langle B_p^\dag\rangle\rangle_t^\mathrm{Coulomb}$ expansion in $\hat{\xi}/\omega_0$. For $p$ being a bound state, $\hat{\lambda}_{000p}=\lambda_D(a_\mathrm{X}/L)^D$ and $\hat{\xi}_{000p}=\xi_DR_\mathrm{X}(a_\mathrm{X}/L)^D$, where $a_\mathrm{X}$ and $R_\mathrm{X}$ are the exciton Bohr radius and Rydberg energy, $D$ the space dimension, $L$ the sample size and $\lambda_D$, $\xi_D$ numerical prefactors of the order of 1. Analytical values of these prefactors for $p=0$ can be found in refs.\ \cite{Phys.Rep.,Pol1}.

For $t$ small, i.e., for $\omega t\ll 1$, we do have $\Delta_t^{(n)}\simeq (it)^n/n$, so that $\langle B_p^\dag\rangle_t^{(3)}$ rises as
\begin{equation}
\langle B_p^\dag\rangle_t^{(3)}\simeq  e^{iE_pt}\Omega_0|\Omega_0|^2\left[2\hat{\lambda}_{000p}
\frac{(it)^3}{3}-\hat{\xi}_{000p}\frac{(it)^4}{4}\right]\ .
\end{equation}

More generally, the recursion relation between the $\Delta_t^{(n)}$'s gives the $t$ dependence of the Pauli part of $\langle B_p^\dag\rangle_t$ as 
\begin{eqnarray}
\Delta_t^{(3)}(\omega_p,-\omega_0,\omega_0)=\hspace{4.5cm}\nonumber\\
\omega_0^{-2}\left[\Delta_t^{(1)}(\omega_p+\omega_0)
+\Delta_t^{(1)}(\omega_p-\omega_0)-2\Delta_t^{(1)}(\omega_p)\right],
\end{eqnarray}
while for the Coulomb part, we find
\begin{eqnarray}
\Delta_t^{(4)}(\omega_p,-\omega_0,-\omega_0,\omega_0)=\omega_0^{-3}\left[3\Delta_t^{(1)}(\omega_p-\omega_0)\right.\nonumber\\
\left.+\Delta_t^{(1)}(\omega_p+\omega_0)-\Delta_t^{(1)}(\omega_p-2\omega_0)-3\Delta_t^{(1)}(\omega_p)\right].
\end{eqnarray}
where $\Delta_t^{(1)}(\omega)$ is essentially a delta function in the large $t$ limit. Fig.1 shows the real parts of $\Delta_t^{(n)}$ for $n=(1,3,4)$, when $p=0$.

\begin{figure*}
\centerline{\scalebox{0.28}{\includegraphics{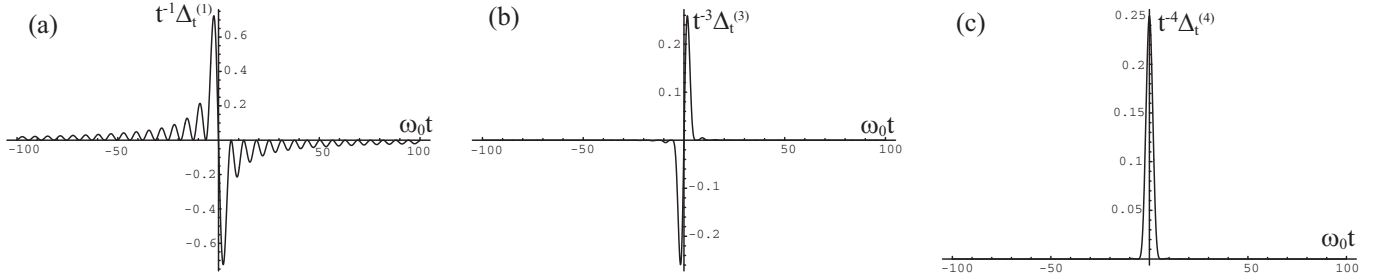}}}
\caption{(a): Real part of $t^{-1}\Delta_t^{(1)}(\omega_0)$, defined in Eq.(9), as a function of $\omega_0t$. (b) and (c): Real parts of $t^{-3}\Delta_t^{(3)}(\omega_0,-\omega_0,\omega_0)$ and $t^{-4}\Delta_t^{(4)}(\omega_0,-\omega_0,-\omega_0,\omega_0)$ defined in Eqs.(22,23), $\Delta_t^{(n)}$ having the same parity as $n$. These curves are more and more picked around $\omega_0=0$ when $t$ or $n$ increases.}
\end{figure*}

\section{State of the art}

Polarization  is usually derived in the free-pair basis by calculating $p_{\v k}=\langle a_{\v k} b_{-\v k}\rangle_t$, through $-i\frac{\partial}{\partial t}p_{\v k}=\langle[(H_{sc}+W_t),a_{\v k} b_{-\v k}]\rangle_t$. In this commutator, the Coulomb part of $H_{sc}$ generates expectation values like
$\langle a^\dag a^\dag aa\rangle_t$, $\langle b^\dag b^\dag bb\rangle_t$ and $\langle a^\dag b^\dag
ba\rangle_t$. These are then cut by Hartree-Fock (HF) procedure, as necessary to get a closed set of differential equations between $p_{\v k}$ and carrier densities,
$f_{\v k}^{(e)}=\langle a_{\v k}^\dag a_{\v k}\rangle_t$ and
$f_{\v k}^{(h)}=\langle b_{\v k}^\dag b_{\v k}\rangle_t$. These equations are then solved numerically using a $4^{th}$ order Runge-Kutta scheme. 

This SBE procedure is presented in details in ref.\cite{Haug1}. It however
has some unpleasant drawbacks, such as divergence in carrier-carrier scattering rates coming from HF approximation, which has to be cut in a ``controlled'' way. Standard HF approximation allows one to recover excitonic poles \cite{Haug1}, but higher-order Coulomb correlations are out of reach. These, including interband electron-hole exchange, can be taken into account, in the case of equilibrium, through the Bethe-Salpeter equation \cite{Sham1966,Hanke1974,Hanke1975,Hanke1980,Onida}, or through nonequilibrium Green functions \cite{Kadanoff,Keldysh}. Both techniques are numerically quite demanding \cite{Onida,Dahlen}. It has recently been found that single-exciton SBE based on time-dependent density functional theory (TDDFT) provide a much simpler approach \cite{Turkowski}. However, they cannot yet account for two-exciton correlations in the single-exciton polarization. In particular, dark excitons with spin $(\pm 2)$ are not reached \cite{Erem} while, evidently, such dark excitons are produced by carrier exchange between two bright excitons \cite{BEC1,BEC2}. Furthermore, the biexciton binding energy cannot be reached from single-exciton SBE. Many important dynamical effects are thus completely lost in these equations. 
Possible improvement can come from dynamics-controlled truncation (DCT) at the two-exciton level in order to obtain biexcitonic correlation effects \cite{Haug1,Axt,Lindberg2}. This DCT method is similar to the BBGKY (Bogoliubov-Born-Green-Kirkwood-Yvon) hierarchy in statistical mechanics for the time evolution of a many-particle system. In any case, since the obtained results rely on heavy numerics, their physical understanding is difficult to grasp because these equations are supposed to contain ``everything'' at once, so that all effects are mixed in the final results, given through numerical curves.

The procedure presented here is far more transparent since it allows us to find analytical expressions for the $t$ dependence of the excitonic polarization, up to any arbitrary order in Coulomb processes and photon field --- this photon field expansion being fully consistent with considering one-pair operators instead of polariton operators.
Our new procedure has the huge advantage to trace back all dynamical effects to these expansion terms. It is in particular possible to distinguish effects arising from Pauli blocking, from effects arising from Coulomb interaction.

\section{Conclusion}

We here develop an alternative approach to semiconductor Bloch equations which allows us to calculate the time evolution of the polarization induced by a laser pulse \emph{analytically}. This approach completely avoids the heavy numerics associated with solving these equations as well as the spurious singularities originating from their necessary truncations. 

Such an analytical solution has been made possible because we here use correlated pairs, i.e., excitons, instead of free pairs, their interaction, responsible for the polarization, being handled through the new composite-exciton many-body theory \cite{Phys.Rep.}. This analytical approach makes the two physical channels producing the polarization completely transparent: fermion exchange and fermion interaction, the former being dominant at large photon detuning. Our approach also deals with the biexcitonic resonance in an easy way, this resonance being uneasy to reach within the usual semiconductor Bloch equation formalism.

\acknowledgments
M.C. acknowledges a one-month invitation by University of Central Florida at Orlando.
M.N.L. acknowledges a one-month invitation by University Pierre et Marie Curie in Paris, as well as support from NSF ECCS-0725514, DARPA/MTO HR0011-08-1-0059, NSF ECCS-0901784, and AFOSR FA9550-09-1-0450.

\section{Appendix}

For readers interested in technical aspects of this approach, we now give the main steps leading to Eqs.(13,15,18).

(a) The Pauli part of $\langle\langle B_p^\dag\rangle\rangle_t$ comes from $\langle\tilde{\psi}_t|D_{ip}|\tilde{\psi}_t\rangle$. By using Eqs.(10,11), this mean value can be written as
\begin{eqnarray}
\langle\langle D_{ip}\rangle\rangle_t=\int_{-\infty}^{+\infty}\frac{dx'}{2i\pi}\int_{-\infty}^{+\infty}\frac{dx}
{(-2i\pi)}\hspace{2cm}\nonumber\\
e^{i(x'-i0_+)t}\,e^{-i(x+i0_+)t}\mathcal{D}_{ip}(x'-i0_+,x+i0_+)\ ,
\end{eqnarray}
where, since $D_{ip}|v\rangle=0$ while $\tilde{H}_{sc}|v\rangle=0$, we do have 
\begin{equation}
\mathcal{D}_{ip}(a',a)=\frac{1}{a'\,a}\sum_{n,n'=0}^{+\infty} \langle\phi_{n'}(a')|D_{ip}|\phi_n(a)\rangle\ ,
\end{equation}
\begin{equation}
|\phi_n(a)\rangle=
\left(\frac{1}{a-\tilde{H}_{sc}}\,\tilde{W}\right)^n\frac{1}{a-\tilde{H}_{sc}}\,\tilde{U}^\dag|v\rangle\ .
\end{equation}
Due to Eqs.(1,2), $\langle v|B_jD_{ip}B_k^\dag|v\rangle$ is simply equal to $2\lambda_{ijkp}$. This gives the quadratic term in photon coupling of $D_{ip}(a',a)$, which comes from $n=n'=0$ in Eq.(25), as
\begin{equation}
\mathcal{D}_{ip}^{(2)}(a',a)=\frac{2}{a'a}\sum_{jk}\frac{\Omega_j \Omega_k^\ast}{(a'+\omega_j)(a+\omega_k)}\,\lambda_{ijkp}\ .
\end{equation}
$\langle\langle D_{ip}\rangle\rangle_t^{(2)}$ follows from the above equations inserted in Eq.(24). Eq.(8) then leads to Eq.(13) for the cubic contribution to $\langle\langle B_p^\dag\rangle\rangle_t$ induced by Pauli blocking. Higher order contributions in photon coupling, which come from $(n,n')\neq (0,0)$ terms in Eq.(25),  can be obtained in the same way. 

(b) The Coulomb part of $\langle\langle B_p^\dag\rangle\rangle_t$ comes from
$\langle\langle V_p^\dag\rangle\rangle_t$ which reads as $\langle\langle D_{ip}\rangle\rangle_t$ in Eqs.(24,25), $D_{ip}$ being simply replaced by $V_p^\dag$.
Since $V_p^\dag$  creates a pair while $V_p^\dag|v\rangle=0$, the lowest order term in photon coupling, obtained for $n=0$ and $n'=1$ instead of $n=n'=0$ as for Eq.(28), then gives
\begin{equation}
\mathcal{V}_p^{(3)}(a',a)=\frac{1}{a'a}\sum_{ijk}\frac{\Omega_i\Omega_j\Omega_k^\ast}{(a'+\omega_j)(a+\omega_k)}\,C_{ijpk}(a')\ .
\end{equation}
Eq.(4) allows us to write
\begin{eqnarray}
C_{ijpk}(a')&=& \langle v|B_iB_j\,\frac{1}{a'-\tilde{H}_{sc}}\,V_p^\dag B_k^\dag|v\rangle\nonumber\\
&=&
\sum_{m,n}M_{ijmn}(a')\xi\left(^{n\ \,k}_{m\ p}\right)\ .
\end{eqnarray}
The exact calculation of
\begin{equation}
M_{ijmn}(a')=\langle v|B_iB_j\frac{1}{a'-\tilde{H}_{sc}}B_m^\dag B_n^\dag|v\rangle\ ,
\end{equation}
requires the knowledge of the whole two-pair eigenstate spectrum.

(i) For materials with well-separated molecular biexciton, this particular two-pair eigenstate controls $M_{ijmn}(a')$ for photons close to the biexciton resonance. When inserted into Eq.(30), the $|XX\rangle$ biexciton state gives
\begin{equation}
M_{ijmn}(a')\simeq\frac{\langle v|B_iB_j|XX\rangle\langle XX|B_m^\dag B_n^\dag|v\rangle}{a'+2\omega-E_{XX}}\ .
\end{equation}
Eq.(29) then leads to
\begin{equation}
C_{ijpk}(a')\simeq \frac{2\xi_{ijkp}^{(XX)}}{a'+2\omega-E_{XX}}\ ,
\end{equation}
with $\xi_{ijkp}^{(XX)}$ given by Eq.(16). Inserting the above result into Eq.(28), we get $\langle\langle V_p\rangle\rangle_t$ through an equation similar to Eq.(24), from which it is easy to find the Coulomb part of the polarization given in Eq.(15).

(ii) $M_{ijmn}(a')$ can also be calculated by iteration. Indeed, Eq.(17) leads to
\begin{eqnarray}
(a'+\omega_i+\omega_j)M_{ijmn}(a')=\hspace{2cm}\nonumber\\
\langle v|B_iB_jB_m^\dag B_n^\dag|v\rangle+\sum_{p,q}\xi\left(^{j\ q}_{i\  p}\right)M_{pqmn}(a')\ ,
\end{eqnarray}
the first term in the above equation being equal to $\delta_{i,m}\delta_{j,n}-\lambda\left(^{j\ \,n}_{i\ m}\right)+(m\leftrightarrow n)$, due to Eqs.(1,2). By keeping this first term only, we get the first term of the $M_{ijmn}(a')$ expansion in Coulomb scattering divided by detuning. Eq.(29) then gives 
\begin{equation}
C_{ijpk}(a')\simeq \frac{2\hat{\xi}_{ijkp}}
{a'+\omega_i+\omega_j}\ ,
\end{equation}
with $\hat{\xi}_{ijkp}$ defined in Eq.(19). The  third-order contribution in photon coupling coming from Coulomb interaction with exciton $p$, as given in Eq.(18), then follows from Eq.(28).


\begin{thebibliography}{0}
\bibitem{unit.transf.} 
This transformation is usually taken symmetrical with respect to electron and hole operators but this is unnecessary.
\end{thebibliography}
\end{document}